\documentclass[pra,twocolumn,showpacs,letterpaper,showpacs,superscriptaddress,nofootinbib]{revtex4-1}
\usepackage{graphicx,amsmath,amssymb,amsfonts,latexsym,color,dcolumn,bm,subfigure}

\begin{document}

\title{Microscale electromagnetic heating in heterogeneous energetic materials based on X-ray CT imaging}

\author{W. J. M. Kort-Kamp}
\affiliation{Center for Nonlinear Studies,  MS B258, Los Alamos National Laboratory, Los Alamos, New Mexico 87545, USA}
\affiliation{Theoretical Division, MS B213, Los Alamos National Laboratory, Los Alamos, New Mexico 87545, USA}
\author{N. L. Cordes}
\affiliation{Materials Science and Technology Division, MS E549, Los Alamos National Laboratory, Los Alamos, New Mexico 87545, USA}
\author{A. Ionita}
\affiliation{Theoretical Division, MS B221, Los Alamos National Laboratory, Los Alamos, New Mexico 87545, USA}
\author{B. B. Glover}
\affiliation{Explosive Science and Shock Physics Division, MS P952, Los Alamos National Laboratory, Los Alamos, New Mexico 87545, USA}
\author{A. L. Higginbotham Duque}
\affiliation{Explosive Science and Shock Physics Division, MS C920, Los Alamos National Laboratory, Los Alamos, New Mexico 87545, USA}
\author{W. L. Perry}
\affiliation{Explosive Science and Shock Physics Division, MS C920, Los Alamos National Laboratory, Los Alamos, New Mexico 87545, USA}
\author{B. M. Patterson}
\affiliation{Materials Science and Technology Division, MS E549, Los Alamos National Laboratory, Los Alamos, New Mexico 87545, USA}
\author{D. A. R. Dalvit}
\affiliation{Theoretical Division, MS B213, Los Alamos National Laboratory, Los Alamos, New Mexico 87545, USA}
\author{D. S. Moore}
\affiliation{Explosive Science and Shock Physics Division, MS P952, Los Alamos National Laboratory, Los Alamos, New Mexico 87545, USA}

\date{ \today}

\begin{abstract}
Electromagnetic stimulation of energetic materials provides a noninvasive and nondestructive tool for detecting and identifying explosives.
We combine structural information based on X-ray computed tomography, experimental dielectric data, and electromagnetic full-wave simulations, to study microscale electromagnetic heating of 
realistic three-dimensional heterogeneous explosives. We analyze the formation of electromagnetic hot spots and thermal gradients in the explosive-binder meso-structures, and compare the heating rate
for various binder systems. 
\end{abstract}


\maketitle

\section{Introduction}

Stand-off detection of explosives via electromagnetic impulses and their resulting spectral responses has been proposed in the literature \cite{Brown2015}. Electromagnetic radiation can also serve as a probe to study the basic physics of light-matter interactions in such systems, which encompass optical, thermal, and chemical processes over a range of spatial and temporal length scales. 
Explosive reactions become self-sustaining (ignition) after a heat source raises the temperature, either locally or globally, to a level where the rate of heat generation by decomposition chemistry becomes appreciable and the heat transport conditions do not allow the heat to dissipate. If the critical conditions are met, the temperature will increase to the combustion temperature and the reaction may spread, depending on other factors. The literature reveals a variety of mechanisms that include acoustic, mechanical, and electromagnetic energy sources that can cause the dynamic deposition of energy into an energetic material. The nature of ignition, or the stimulation of chemistry, then depends on the specific physical interactions (molecular, atomic, electronic, mechanical, etc.) and their spatial character. Interesting behavior arises in heterogeneous materials having contrast in these interactions within the microstructure.
In the case of electromagnetic stimulation of heterogeneous energetic materials, hot spots can form in regions of large contrast of the complex electrical permittivity of the constituent materials, leading to enhanced EM fields and large micro- to meso-scale temperature gradients \cite{Perry2014}.
Localization of electromagnetic energy under sub-initiation threshold stimulation can potentially generate local temperature increases sufficient to release some of the chemical energy stored in explosive molecules, subsequently leading to alternative signatures for detection based on nonlinear coupling of light to energetic materials.

Recent experimental and theoretical works on microwave heating of energetic materials have aimed at understanding the multi physics problem of electromagnetic energy localization, thermal hot spot generation, and induced chemistry leading to deflagration or detonation
\cite{Perry2008,Curling2009,Perry2011,Daily2013,Perry2014}. In this highly non-equilibrium phenomena involving electromagnetic dispersion, thermal gradients and complex chemistry, it is in general very challenging, if not impossible, to experimentally measure temperature distributions, hot spots, and phase transition dynamics. Modeling and simulation tools are therefore a valuable route towards understanding and predicting the behavior of heterogeneous energetic materials under electromagnetic stimuli. Previous works along this direction have used simple semi-analytical models based on homogenization theory \cite{Perry2008}, or continuum models of idealized mesoscale geometries to mimic real complex structures \cite{Perry2011,Perry2014}. Although useful, these approaches cannot unequivocally determine the spatial and temporal behavior of hot spots in real heterogeneous energetic materials.

A useful method for obtaining three dimensional (3D) models of real structures is through nondestructive micro-scale X-ray computed tomography (CT) \cite{Stock2008}. This imaging technique can be used to acquire spatially-resolved 3D information of several types of materials \cite{Patterson2010,Patterson2013,Cordes2015,McIntosh2015}.
CT images have been used to model physical processes within a wide range of materials, such as deformations of cellular solids \cite{Maire2003,Petit2013,Patterson2016} and highly porous materials \cite{Maire2012}, biomechanics of living and fossil organisms \cite{Rayfield2007}, biomechanics in dental research \cite{Swain2009}, damage behavior in asphalt materials \cite{Dai2005}, and in assessing fruit quality \cite{Mebatsion2008}. While there have been studies which have utilized CT to visualize \cite{Tringe2013} and quantify internal void structures of polymer bonded  explosives (PBX) 
\cite{Thompson2010,Willey2015} and mock sugar \cite{Forsberg2009} systems (sugar is often used as a surrogate material for high explosives), the utilization of CT-generated models for finite element analysis of PBX has not been previously reported. 

Here, we perform the first study of electromagnetic stimulation of energetic materials using X-ray computed tomographic (CT) 3D data of real heterogeneous explosives based on HMX (octahydro-1,3,5,7-tetranitro-1,3,5,7-tetrazocine) crystals, a very common
secondary high explosive and propellant. We measure the complex spatial distribution of explosive crystals within a binder matrix, and use our experimentally determined frequency- and temperature-dependent dielectric data to perform full-wave simulations of electromagnetic microwave heating of energetic materials. We analyze the formation of 
electromagnetic hot spots and thermal gradients within the meso-structure.

The paper is organized as follows. In Section II we describe our CT measurements of plastic bonded explosives, and the necessary image processing to obtain quality 3D images of the heterogenous structure amenable to full-wave electromagnetic simulations. Measurements of  frequency and temperature dependence of the complex permittivity of energetic materials are contained in Section III. Section IV contains the numerical simulations of the CT data using commercial multi physics (COMSOL) numerical solver. We also compare our simulation results with a simple theoretical model based on effective medium theory. Finally, Section V contains an outlook for future work, e.g., the addition of possible nonlinear effects, the inclusion of complex chemistry to describe various phases along the decomposition kinetics, and the extension to other interesting frequency regimes  (THz) with potential applications in explosive detection.


\section{CT data}

Micro-scale X-ray computed tomography is achieved by acquiring radiographs of a sample as a function of sample rotation (typically $180^o$). Reconstructing the radiographs yields a digital representation of the sample in 3D, revealing both surface and subsurface features. Contrast within the tomogram (i.e., 3D image data set) is based on the total X-ray attenuation of a material. Therefore, tomograms of materials which are heterogeneous in composition and density exhibit high contrast, suitable for image processing, analysis, and quantification. Tomograms are typically segmented (i.e., binarized) for each material phase for analysis. The segmented data sets are also suitable to be converted into a 3D mesh of tetrahedral elements and fed to a finite element modeling program for numerical calculations. 

One difficulty in generating models based on CT-generated structures with some heterogeneous energetic materials, such as PBX 9501 ($95\%$ HMX, $2.5\%$ Estane, $2.5 \%$ BDNPA/F nitroplasticizer), is the insufficient contrast within the resulting tomograms (e.g., see Fig. 2 of \cite{Rayfield2007} and Fig. 5 of \cite{Dai2005}). This in turn greatly hinders separate grayscale-based segmentations of the two phases and prevents the generation of 3D models for meshing. In this study, the Estane binder normally found in PBX 9501 has been replaced by hydroxyl-terminated polybutadiene (HTPB), which significantly enhances the contrast between the HMX phase and binder phase in the tomogram and allows for separate grayscale-based segmentations of these two phases.  With the HTPB phase acting as a spatial surrogate for Estane, known Estane properties can be assigned to the HTPB volume in the finite element modeling. The PBX used in this study consisted of, by weight, 88\% HMX, 5.4\% hydroxyl-terminated polybutadiene (HTPB) as the binder, 5.4\% dioctyladipate, and 0.5 \% methylene biphenyl diisocyanate (MDI). 
Samples were combined on a mass balance and mixed by hand. The material was then cored into 5-mm cylinders using a brass cork borer.
\begin{figure} 
\includegraphics[width=3.3in]{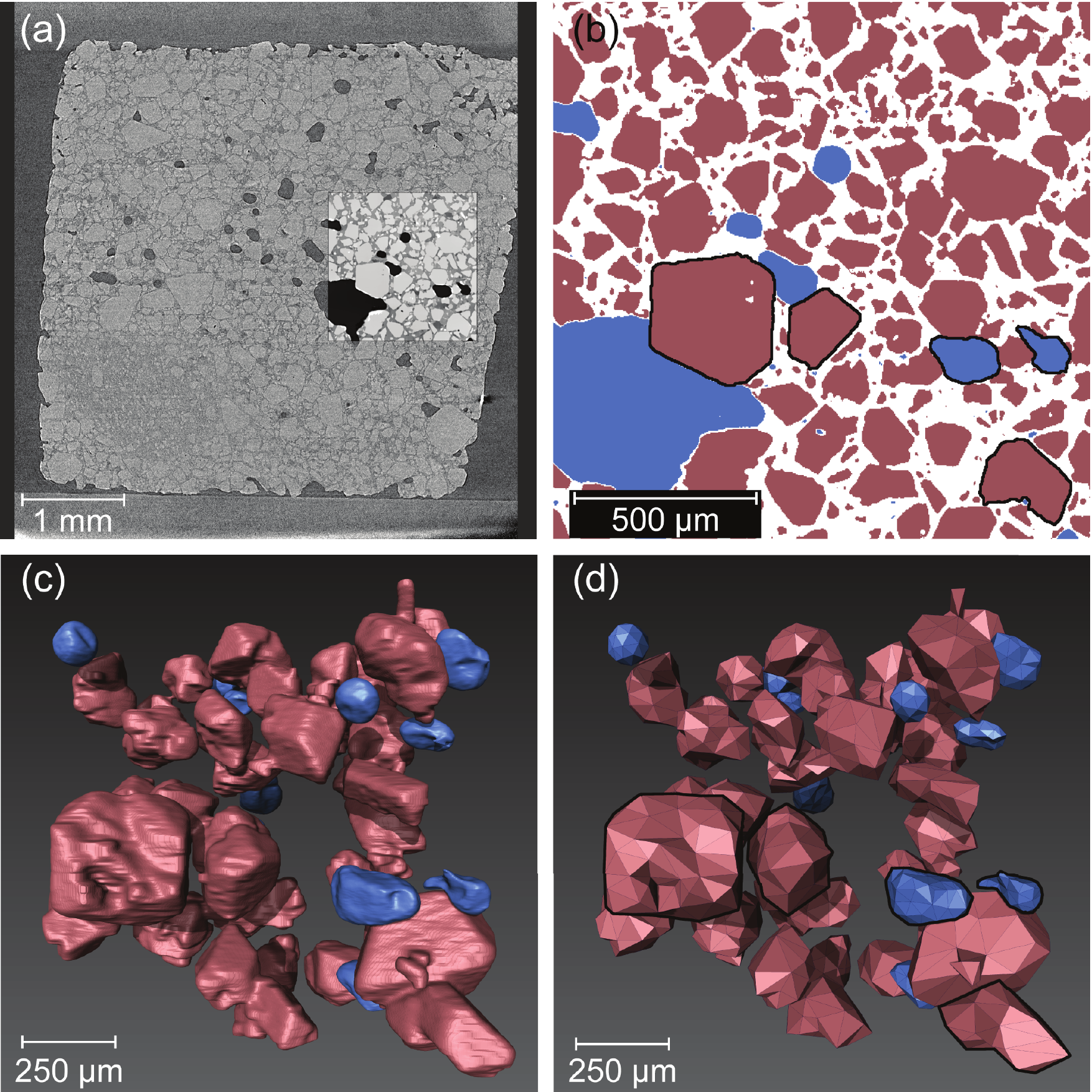}  
\caption{
(Color online) (a) Reconstructed slice of PBX 
(b) Segmented cropped volume [highlighted in (a)] after median smoothing and diffusion-based smoothing.
Blue objects are voids, the white region is the HTPB binder, and red objects are HMX crystals. (c) High-quality surface rendering of volume in (b) after image processing (see text). (d) Simplified surface rendering of volume in (c) used in our COMSOL simulations.
The black outlined objects correspond to the black outlined objects in (b).} 
\label{Fig1}
\end{figure}

A micro-scale X-ray computed tomogram of the PBX sample was collected using a Carl Zeiss X-ray Microscopy, Inc., (Pleasanton, CA) MicroXCT X-ray microscope operating with a W anode X-ray source in a cone beam geometry and a 4X objective lens. Operating conditions were: applied voltage of 40 kVp and applied current of 250 $\mu$A; camera binning of 1; 30 s exposure per radiograph. The sample was rotated $186^o$ with $0.093^o$ per radiograph resulting in 2001 radiographs. Tomogram reconstruction was performed using Xradia XMReconstructor software. The resulting tomogram (Fig. 1a) exhibited an isotropic voxel size of 2.76 $\mu$m. 
Post-reconstruction image processing was performed using Avizo (Version 9.0, FEI Visualization Sciences Group, Burlington, MA). 
To reduce noise, a median smoothing filter was applied to the tomogram.
Before further image processing, a 538 pixel x 538 pixel x 538 pixel volume was cropped from the tomogram (highlighted region in Fig. 1a). To further reduce noise and enhance contrast within the tomogram, a diffusion-based smoothing filter was applied.
Segmentation of the cropped volume was performed 
resulting in a ternary data set consisting of three segmented phases (HMX, HTPB binder, and voids) (Fig. 1b).
 
Further image processing was necessary to generate surface files suitable for importing into COMSOL:  
the segmented data set was separated into each individual phase, a sieve filter was used to remove small volumes 
($< 150\ \mu$m$^3$), objects which impinged on the six borders of the data set were removed, and holes within the crystals were also removed. 
The resulting data sets yielded 33 HMX crystals and 15 void regions. Surface files 
of the processed images were generated, and then reduced from $\simeq 2.0 \times 10^6$ HMX and $\simeq 1.0 \times 10^6$ 
void faces (Fig. 1c) to $2.6 \times 10^3$ HMX and $1.4 \times 10^3$ void faces (Fig. 1d). Finally, the simplified surface files were processed (e.g., fix intersections, fix small dihedral angles, etc.) before being exported for mesh generation in COMSOL.


\begin{figure} [t]
\vspace{0.5cm}
\includegraphics[width=3.3in]{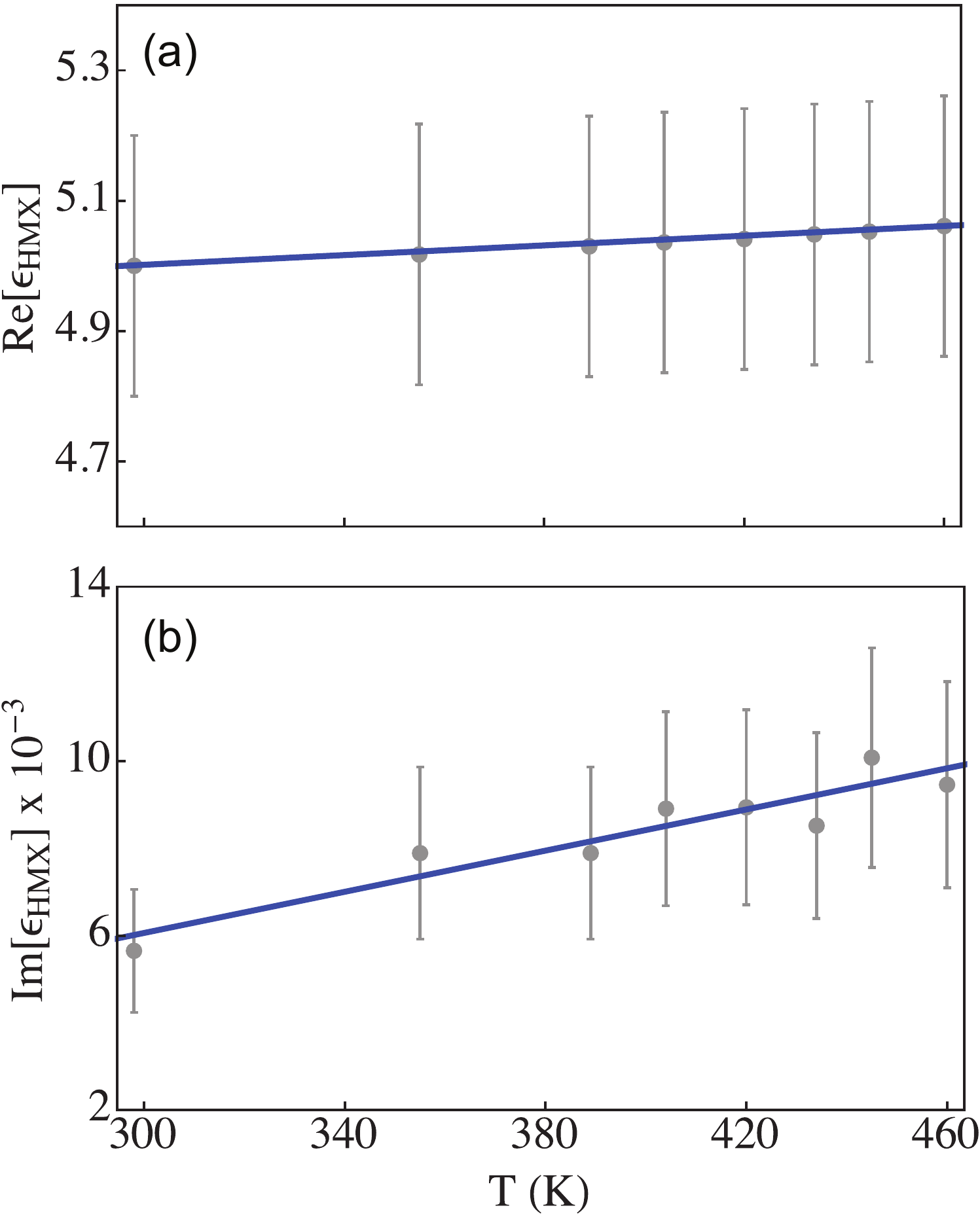}  
\caption{
(Color online) Real (top) and imaginary (bottom) permittivity as a function of temperature for neat HMX. The experimentally measured values have been corrected to reflect data at TMD. Although it appears the data has little to no dependence on temperature within the precision error of our instrument, the data was fit using a linear approximation for the purpose of obtaining input parameters for the electromagnetic heat transfer simulations.} 
\label{Fig2}
\end{figure}

\section{Dielectric and thermal data}

In this section we describe the dielectric and thermal properties of HMX and various binders that will be required to perform the electromagnetic heat transfer simulations in heterogeneous PBX systems. 

The measurement of the dielectric properties of secondary high explosives in the microwave frequency range has been recently performed \cite{Duque2014}. Using a combination of a resonant cavity perturbation technique for determining the room temperature complex dielectric constant at discrete frequencies and a broadband open circuit method (1-18 GHz), the frequency and density dependence of the dielectric constant of various energetic materials at room temperature were reported. It was determined that in this particular frequency range, the data is constant and largely independent of frequency.

Using the same broadband open circuit technique as in \cite{Duque2014}, we studied the temperature dependence from 295-460 K of the permittivity of HMX at 1-18 GHz. The metal coaxial fixture was heated in a vertical tube furnace (both empty and with the sample present) so that the response of the fixture could be subtracted out at elevated temperatures, thus isolating the material response. The temperature of the sample and surrounding fixture were continuously monitored using two independent thermocouples, and was allowed to equilibrate for 30 minutes at each target temperature before data was acquired. The results are shown in Figure 2 and have been corrected to represent data at theoretical maximum density (TMD) $\rho=1.91 \; {\rm g/cm^3}$; the single points represent the average of the data at 1-18 GHz since there is no frequency dependence in this range for these low-loss, dielectric materials. The error bars are a liberal interpretation based on the precision of measurements made at discrete frequencies (described in \cite{Duque2014}); this was used because the temperature dependent data was corrected for density based on the corrected real permittivity room temperature value for HMX at TMD calculated in \cite{Duque2014}.

We determined that for neat HMX, there does not appear to be a temperature dependence of the dielectric constant within the precision error of our instrument that may be attributed to temperature effects alone. The reason for the weak dependence on temperature is that the available polarization and resonance mechanisms (electronic, distortion, space charge, atomic, etc.) themselves do not change significantly with temperature for this molecule. Density reduction will have an effect on the real component, but over the temperature range investigated we expect about a $2\%$
change in crystal density and a commensurate change in the permittivity. The effect of density change on the imaginary component should be negligible. However, the slight linear increase of the real permittivity as a function of temperature (likely due to the density reduction as the sample was heated) was fit for the purpose of obtaining input parameters for the electromagnetic heat transfer simulations. The following form was used:
\begin{equation}
\epsilon_{\rm HMX}(T) = \epsilon_{\rm HMX}(T_a) + \eta(T-T_a),
\end{equation}
where $\epsilon_{\rm HMX}(T)$ is the permittivity as a function of temperature $T$, $T_a$ is room temperature, 
and $\eta=\eta_R + i \eta_I$ is a complex slope. 
We obtain $\eta_R=3.72\times 10^ {-4}$ K$^{-1}$ and $\eta_I = 2.36\times 10^{-5}$ K$^{-1}$.

Broadband room temperature permittivity data for various PBX systems was reported in \cite{Duque2014}, including PBX 9501 
and PBXN-5 ($95\%$ HMX, $5\%$ Viton A). To our knowledge, broadband and temperature-dependent dielectric property data of the corresponding binders alone is unfortunately not available. Because our primary goal in this work is to perform full-wave simulations of heterogeneous energetic materials at the microscale using the individual components, we will use the room temperature permittivity of the binders at selected frequencies (2.5 GHz and 13.3 GHz) reported in \cite{Duque2014}. These materials also appear to display constant response as a function of frequency; for example the measured permittivity of Estane:BDNPA/F (1:1) is $\epsilon= 4.0 + i 0.38$, and for Viton A is $\epsilon = 2.5 + i 0.09$, at both of these frequencies. To our knowledge, the temperature dependency of the binder permittivity has not been reported in the literature. Therefore in our COMSOL simulations we will assume that the permittivity of the binder is weakly dependent on temperature in the range we will study. Once temperature-dependent dielectric property data for the binders becomes available in the future, it can be easily incorporated into the numerical simulations described in the next section.

Finally, we briefly quote thermal data for our PBX systems. The temperature-dependent specific heat $C_p(T)$ and thermal conductivity $k(T)$ of the HMX crystals 
in their $\beta$-phase are modeled as linearly dependent on temperature, following \cite{Henson2009}. Namely, $C_p(T)=C_1 + C_2 T$ 
(with $C_1=248.9 \, {\rm J/kg \, K}$ and $C_2=2.616 \, {\rm J/kg \, K}^2$) and $k(T)=k_1 + k_2 T$ 
(with $k_1=0.4368 \, {\rm W/m \, K}$ and $k_2=-4.44 \times 10^{-4} \, {\rm W/m \, K}^2$). The density is taken as $\rho=1910 \; {\rm kg/m}^3$ and assumed to be independent of temperature in the range considered. For the binders, temperature dependency of their specific heat and thermal conductivity is not available in the literature. We will assume the following constant values: For Estane-BDNPA/F(1:1), $\rho= 1200\, {\rm kg/m}^3$, $C_p= 1500$ J /kg K  and $k= 0.25 $ W/m K. For Viton, $\rho= 1800 \, {\rm kg/m}^3$, $C_p= 1464$ J /kg K and $k= 0.226$ W/m K \cite{Dobratz1981}.

\begin{figure}[t]
\includegraphics[width=3in]{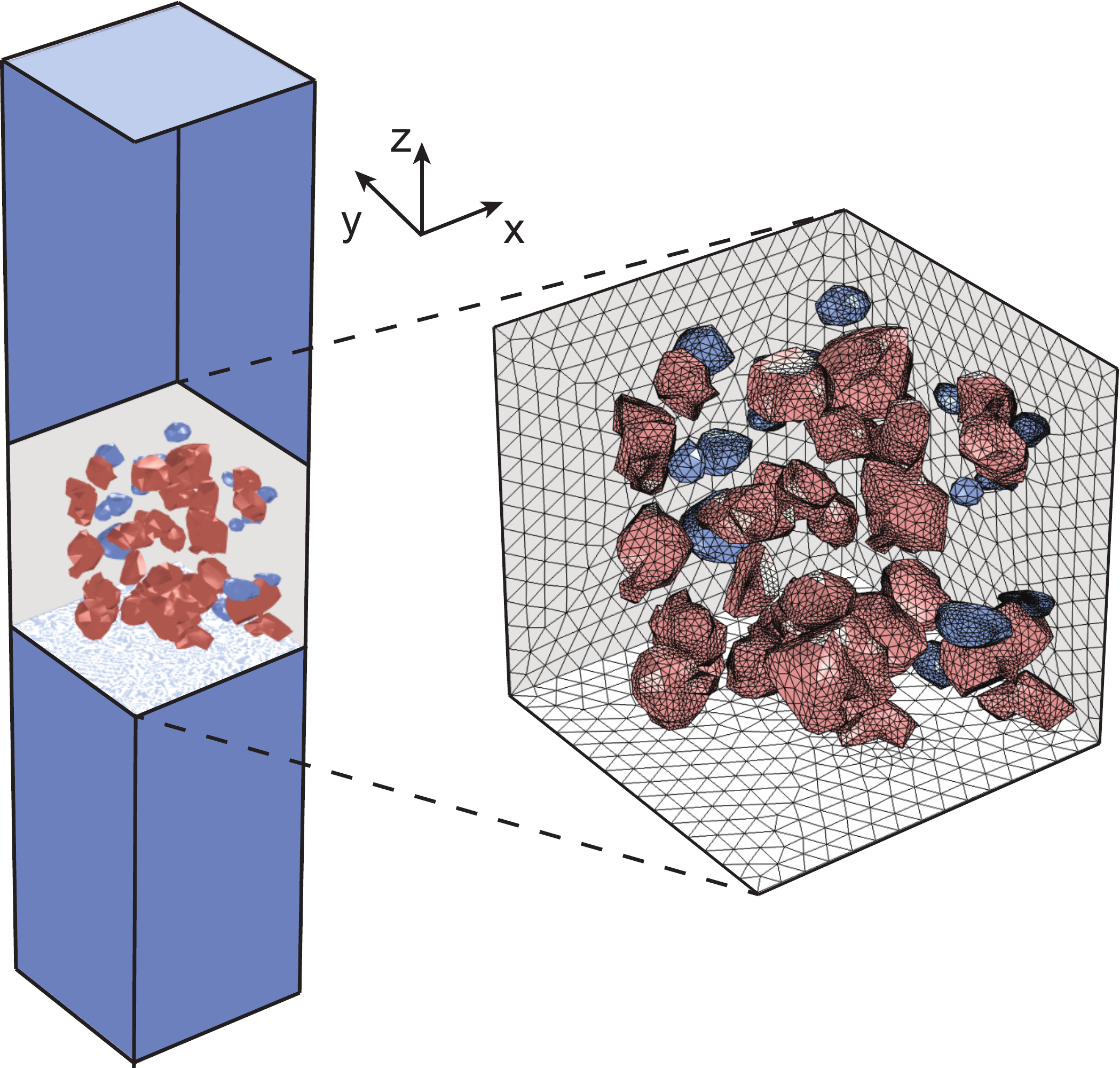}  
\caption{(Color online) Geometry studied to compute the electromagnetic-thermal response of heterogeneous energetic materials using CT data.
A cubic unit cell of size 1.5 mm is periodically replicated in the x-y plane to form an infinite slab, which is immersed in air. 
The unit cell corresponds to HMX crystals (red) and air voids (blue) embedded in a binder slab (gray), and corresponds to the same structure shown in Fig. 1d.
An incident electromagnetic wave impinges at normal incidence on the slab from the $z>0$ air region. 
} 
\label{Fig3}
\end{figure}


\begin{figure*}[t]
\includegraphics[width=7in]{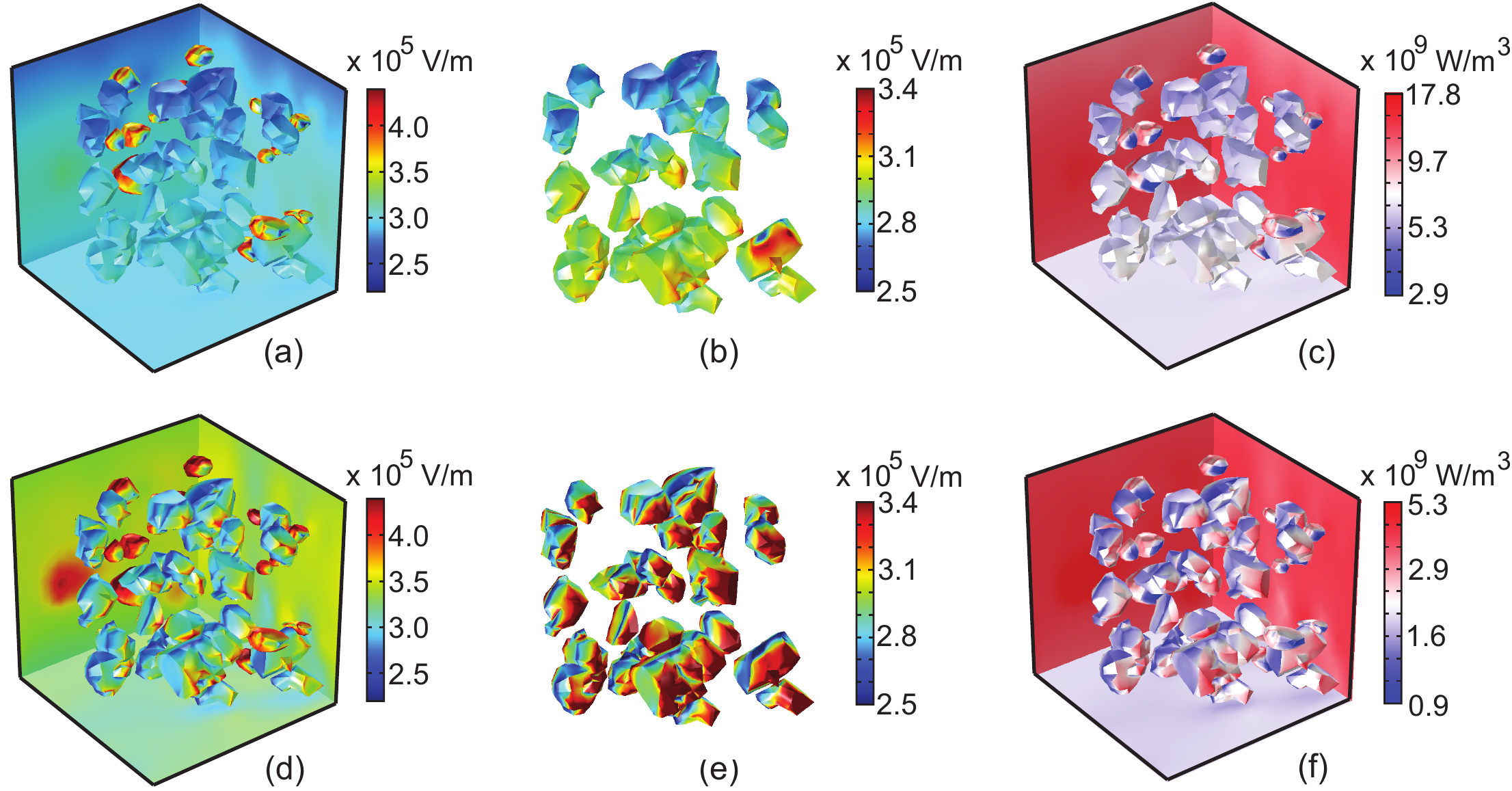}  
\caption{(Color online) Electric field distribution and EM power dissipated in the PBX slab based on Estane-BDNPA/F(1:1) binder (top plots) and Viton A binder (bottom plots).
The electric field amplitude distribution in the slab and only the HMX crystals are presented in (a) and (b), respectively. Panel (c) shows the total power dissipation density profile in the energetic material. Panels (d-f) show the corresponding results for the case using Viton A.}
\label{Fig4}
\end{figure*}


\section{Modeling and simulations}

In order to determine the electromagnetic response of the system it is necessary to solve the coupled Maxwell  and heat transfer equations. Since the characteristic time scales of the electromagnetic field are much shorter than those of the electromagnetic response of the materials, one can perform a separation of time scales and first solve Maxwell equations for fixed materials properties. In this approximation, the frequency $\omega$ of an input plane wave is conserved, and the electromagnetic field satisfies the Helmholtz equation 
\begin{equation}
\nabla^2{\bf E}({\bf r}, \omega) + \epsilon({\bf r}, \omega; T) \dfrac{\omega^2}{c^2} {\bf E}({\bf r}, \omega)  = 0 ,
\label{FieldEquation}
\end{equation}
where the permittivity of the material slowly depends on time through the temperature. For our heterogeneous material, $\epsilon=1$ in the air regions, $\epsilon=\epsilon_{\rm HMX}$ within the HMX crystals, and 
$\epsilon=\epsilon_{\rm binder}$ in the binder. The temperature distribution $T({\bf r}, t)$ is governed by the heat equation,
\begin{equation}
\rho({\bf r})C_p({\bf r}; T) \dfrac{\partial T}{\partial t} = k({\bf r}; T) \nabla^2T + Q({\bf r}; T) .
\label{TemperatureEquation}
\end{equation}
Here, $\rho({\bf r})$ is the density, $ k({\bf r}; T)$ is the thermal conductivity, $C_p({\bf r}; T)$ is the heat capacity, and $Q({\bf r}; T)$ is a heat source. Note that in our numerical simulations we will not consider heat transfer through radiation or convection (the latter is potentially relevant in the air inclusions). In the air regions $Q=0$, in the binder $Q=(\omega/2) {\rm Im}[\epsilon_{\rm binder}(\omega)] |{\bf E}({\bf r},\omega)|^2$ is the absorbed electromagnetic power density, and in the HMX crystals $Q$ is given by the sum of the electromagnetic power density and the chemical thermal source, 
$Q=(\omega/2) {\rm Im}[\epsilon_{\rm HMX}(\omega)] |{\bf E}({\bf r},\omega)|^2 + Q_{\rm chem}$. For the temperature variations we will study in this work, phase transition in HMX are negligible because we are not heating past 160 $^o$C where the $\beta$-$\delta$ phase transition begins 
\cite{Cady1961}. We then model 
$Q_{\rm chem}$ by a simple Arrhenius model
$Q_{\rm chem}=\rho \Delta h Z e^{-E_r/R T}$, where $\Delta h=6.2 \times 10^6$J/kg is the reaction enthalpy, $Z=5 \times 10^{19} \, {\rm s}^{-1}$, $E_r=220.5$ kJ/mol is the activation energy, and 
$R=8.314$ J/mol K is the universal gas constant. Furthermore, for simplicity we will neglect electro-mechanical and thermo-mechanical couplings, e.g. thermal expansion. 

We use COMSOL Multiphysics to perform the coupled electromagnetic-thermal simulation. The geometry is a slab of thickness $1.5$ mm with unit cell given in fig. 3, which corresponds to the same CT structure shown in fig. 1d. As discussed in Section II, in the simulations we use a surrogate binder (either Estane-BDNPA/F(1:1)  or Viton A). We use periodic boundary conditions in the $x-y$ plane, both for the EM and thermal fields. A  $y$-polarized plane wave of frequency $\omega/2 \pi =13.3$ GHz and intensity $I=2.2 \times 10^8 \; {\rm W/m^2}$ impinges normal to the slab along the $-z$ direction. We assume that initially the slab is at thermal equilibrium at temperature $T({\bf r}, t=0)=293$ K. 

We first consider the spatial distributions of the electric field and electromagnetic dissipated power within the PBX heterogeneous structure, and compare the results for the two binders under study. The top row in fig. 4 corresponds to Estane-BDNPA/F(1:1) as the binder material in the PBX system, and bottom row corresponds to a PBX based on Viton A.
Fig. 4a we show the electric field amplitude distribution. Note that since the temperature increase within the material is less than 150 K, the (small) variation with temperature of $\epsilon_{\rm HMX}$ (see Fig. 2) is not sufficient to produce any significant changes in the EM field distribution in time. Since the real part of the permittivity of HMX is similar to that of Estane-BDNPA/F(1:1) and the displacement vector ${\bf D}=\epsilon {\bf E}$ must be continuous across material interfaces, there is not much contrast in the amplitude of the electric field on the HMX/binder boundaries. On the other hand, the field jumps by a factor of $\sim 4$ across the air/binder interfaces due to their dielectric contrast. In order to better see the field distribution in the HMX crystal, we show in fig. 4b only the HMX crystals. The overall amplitude distribution of the field, weak to strong from top to bottom, reflects the spatial variation of the incident GHz field. On top of this background, there are local spatial variations where the field is concentrated around the edges of the HMX crystal forming EM hot spots. Fig. 4c shows the power dissipation density in the slab. It is mainly concentrated in the binder due to its larger imaginary permittivity, and even within the binder there is more dissipation close to the air voids since $Q$ is proportional to the square of the field. When Viton A is used as binder, there is a larger variation of the electric field at HMX-binder interfaces compared to the previous case 
(compare Figs. 4a and 4d), due to the fact that HMX and Viton A have more dissimilar real permittivities. The presence of hot spots at the edges of HMX crystals is clearly evinced in Fig. 4d. Fig. 4e and 4f show the corresponding field distributions in the HMX crystals and EM power dissipated in the PBX system based on Viton-A.

We now consider the thermal distribution in the energetic materials based on the two binders under consideration.
Fig. 4a reports the spatial distribution of temperature after 15 ms for Estane-BDNPA/F based PBX.
As seen in the figure, the regions around the air voids heat faster for the above discussed reasons, and the smaller the voids the faster they thermalize with the surrounding binder. Stronger increased local temperatures can be observed around the HMX crystal faces due to the EM hot spots. Fig. 4b shows the temperature distribution on a vertical slice at middle of the slab, $x=750\ \mu$m. 
Fig. 4c and 4d show the corresponding temperature distributions for the case of Viton A based PBX. 
Similarly to the Estane-BDNPA/F(1:1) case, the binder heats up faster than the HMX inclusions since ${\rm Im}[\epsilon_{\rm binder}]>{\rm Im}[\epsilon_{\rm HMX}]$. Besides, note that higher temperatures in HMX regions occur at the same locations where the EM field hot spots are excited. Finally, Fig. 4e shows the average temperature in the slab as a function of time for the two considered binder materials. The smaller variation of  temperature in the case of Viton A is related to the fact that ${\rm Im}[\epsilon_{\rm Estane-BDNPA/F}]\gg{\rm Im}[\epsilon_{\rm Viton}]$.


\begin{widetext}

\begin{figure}[t]
\includegraphics[width=7in]{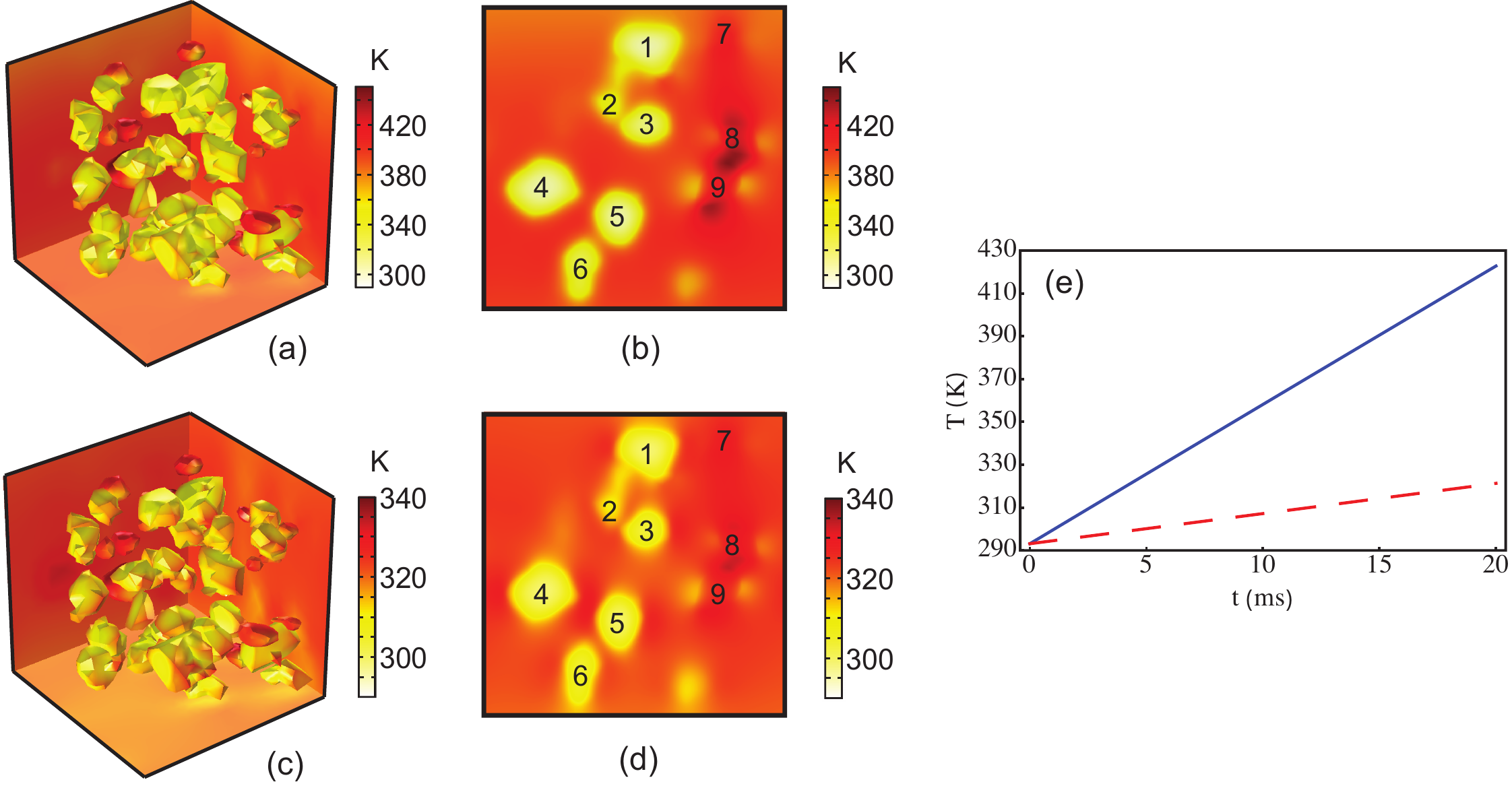}  
\caption{(Color online) Temperature distribution in the PBX slab based on Estane-BDNPA/F(1:1) or Viton A binders.
(a) depicts the temperature field distribution in the slab based on Estane-BDNPA/F binder, and (b) on a vertical slice at $x=750\ \mu$m, respectively, after 15 ms. Rgions 1-6 indicate the position of HMX crystals and regions 7-9 of air voids present in the selected slice. (c-d) show the corresponding results for the Viton-A based PBX system. Panel (e) shows the average temperature in the slab as a function of time for the structure with Estane-BDNPA/F(1:1) (solid) and with Viton A  (dashed).} 
\label{Fig5}
\end{figure}
\end{widetext}

\section{Conclusions}

We have performed the first full-wave simulations of coupled electromagnetics and thermal transport in three-dimensional heterogeneous PBX explosives based on 
our experimentally measured structural and dielectric properties. Our modeling and simulation tools represent a basis to understand and predict
the behavior of heterogeneous energetic materials under electromagnetic stimuli. Our proof-of-principle study is a first step towards a more complete study of light-matter interactions in heterogeneous
explosives. In this work we have limited the interaction times and field intensities so that chemical reactions are not triggered. Future extensions of this work will include chemical decomposition of the explosive
molecular crystals all the way to ignition, effects due to electro-thermo-mechanical couplings, and the possibility of nonlinear electromagnetic effects in energetic materials \cite{Wood2015}, among others.


\section{Acknowledgements}

We thank Virginia Manner and Bryce Tappan of LANL for providing PBX samples. We are grateful to the LANL LDRD program for financial support.


\end{document}